\documentclass[conference]{IEEEtran}
\usepackage{amsmath, amsthm, amssymb, mathtools, abraces, pifont, paralist, relsize, hyperref, xcolor}

\newtheorem{theorem}{Theorem}

\newtheorem{corollary}[theorem]{Corollary}
\newtheorem{lemma}[theorem]{Lemma}
  \theoremstyle{definition}

  \theoremstyle{remark}

\newtheorem{example}{Example}

\newcommand{ \bs }{ \boldsymbol }

\newcommand{ \und }{ \underline }
\newcommand{ \pid }{ \pi^\textnormal{id} }
\newcommand{ \Sout }{ S^{^\mathsmaller{\rightarrow}}\! }
\newcommand{ \SSout }{ \bs{S^{^\mathsmaller{\rightarrow}}}\! }
\newcommand{ \Sin }{ S^{^\mathsmaller{\leftarrow}}\! }
\newcommand{ \SSin }{ \bs{S^{^\mathsmaller{\leftarrow}}}\! }
\newcommand{ \Srev }{ S^{^\mathsmaller{\leftrightarrow}}\! }

\newcommand{ \Smout }{ S_\textsc{m}^{^\mathsmaller{\rightarrow}}\! }
\newcommand{ \SSmout }{ \boldsymbol{S_\textsc{m}^{^\mathsmaller{\rightarrow}}}\! }
\newcommand{ \Smin }{ S_\textsc{m}^{^\mathsmaller{\leftarrow}}\! }

\newcommand{ \Smrev }{ S_\textsc{m}^{^\mathsmaller{\leftrightarrow}}\! }

\newcommand{ \defeq }{ \coloneqq }

\newcommand{\myqed}{\hfill $\blacktriangle$%
                   }

\begin{document}

\title{Some Enumeration Problems in the Duplication-Loss Model of Genome Rearrangement}

\author{
\IEEEauthorblockN{Mladen~Kova\v{c}evi\'{c}, Sanja~Brdar, and Vladimir~Crnojevi\'{c}}
\IEEEauthorblockA{BioSense Institute,
                  University of Novi Sad, 21000 Novi Sad, Serbia\\
                  Emails: \{kmladen, brdars, crnojevic\}@uns.ac.rs}
}

\maketitle

\begin{abstract}
Tandem-duplication-random-loss (TDRL) is an important genome rearrangement
operation studied in evolutionary biology.
This paper investigates some of the formal properties of TDRL operations
on the symmetric group (the space of permutations over an $ \bs{n} $-set).
In particular, the cardinality of ``balls'' of radius one in the TDRL metric,
as well as the cardinality of the maximum intersection of two such balls,
are determined.
The corresponding problems for the so-called mirror (or palindromic) TDRL
rearrangement operations are also solved.
The results represent an initial step in the study of error correction and
reconstruction problems in this context, and are of potential interest in
DNA-based data storage applications.%
\end{abstract}

\IEEEpeerreviewmaketitle

\section{Introduction}
\label{sec:introduction}

The study of genome rearrangements in evolutionary biology is a rich source of
mathematical and algorithmic problems that, apart from their relevance for the
field they originated in, are also interesting in their own right \cite{fertin, pevzner}.
In the present paper, we are concerned with the so-called tandem-duplication-random-loss
(TDRL) model of genome rearrangement, which is of importance in the study of gene
order evolution in mitochondrial genomes \cite{bernt2, mauro}.
Specifically, we focus on the combinatorial questions of finding the cardinalities
of balls and intersections of balls in this context, questions that are
important primarily from a coding theoretic viewpoint, and in particular for
error correction and reconstruction problems.
Our results are of possible interest in DNA-based data storage applications \cite{yazdi}.
Namely, in settings where information is being stored in the form of DNA molecules
(or pools thereof), the naturally occurring mutations and rearrangement operations
represent the ``noise'', and methods of dealing with this noise are therefore essential
for reliable data recovery.

Combinatorial problems inspired by the TDRL rearrangement model have been studied
previously in several works;
see, e.g., \cite{bernt, bouvel+pergola, bouvel+rossin, chaudhuri, hartmann}.

\subsection*{Notation and Terminology}

For our purposes, genome can be modeled as a \emph{permutation} on the set
$ \{1, 2, \ldots, n\} $ \cite{chaudhuri}.
The set of all permutations over $ \{1, 2, \ldots, n\} $ is denoted by $ \Pi(n) $.
Each permutation $ \pi \in \Pi(n) $ is regarded simply as a sequence $ (\pi_1, \pi_2, \ldots, \pi_n) $,
where $ \{\pi_1, \pi_2, \ldots, \pi_n\} = \{1, 2, \ldots, n\} $, and thus the elements
of $ \Pi(n) $ will sometimes be referred to as sequences.
The identity permutation is denoted by $ \pid(n) \defeq (1, 2, \ldots, n) $,
or by $ \pid $ if the length $ n $ is understood from the context.
We say that $ (\pi_{i_1}, \ldots, \pi_{i_m}) $, where $ 1 \leq i_1 < \cdots < i_m \leq n $,
is a subsequence of length $ m $ of the sequence $ (\pi_1, \pi_2, \ldots, \pi_n) $.

\pagebreak
\section{TDRL Permutations}

A TDRL operation on a sequence $ \pi \in \Pi(n) $ is a duplication of the entire
sequence $ \pi $, followed by a deletion of one of the two copies of each of the
symbols.
Thus, each TDRL operation is a permutation of the coordinates of $ \pi $, and the
result is another sequence from $ \Pi(n) $.

\begin{example}
\label{ex:tdrl}
An example of a TDRL operation on $ \pid(5) $ is the following:
\begin{subequations}
\label{eq:tdrl}
\begin{align}
\label{eq:tdrl5}
  &1 \ \und{2} \ \und{3} \ 4 \ \und{5} \ \aoverbrace[L1R]{\und{1} \ 2 \ 3 \ \und{4} \ 5}
	\quad \longrightarrow \quad  2 \ 3 \ 5 \ 1 \ 4  \\
\label{eq:bintdrl}
  &0 \ 1 \ 1 \ 0 \ 1
\end{align}
\end{subequations}
In \eqref{eq:tdrl5}, the duplicate of the original sequence is overbraced,
and the symbols that are \emph{not} deleted are underlined.
\myqed
\end{example}

By definition, the symbols that are deleted from the first copy of $ \pi $ are not
deleted from the second copy, and vice versa.
Therefore, a TDRL operation can be specified by a binary pattern indicating the
symbols that are not deleted from the first copy of a given sequence, as illustrated
in \eqref{eq:bintdrl}.
We will use this binary representation throughout the paper.

Another way to think of a TDRL operation on $ \pi $ is as a partition of $ \pi $
into two of its subsequences which are then concatenated.
For example, in \eqref{eq:tdrl}, $ \pid(5) $ is partitioned into $ (2, 3, 5) $ and
$ (1, 4) $, and the final result is $ (2, 3, 5, 1, 4) $.

If a sequence $ \rho $ is the result of applying a TDRL operation on $ \pi $,
we write $ \pi \to \rho $, and we define
$ \SSout(\pi) \defeq \{ \rho : \pi \to \rho \} $ and
$ \SSin(\pi) \defeq \{ \rho : \rho \to \pi \} $.
The set $ \SSout(\pid) $ is illustrated in Table \ref{tab:tdrl}.

{
\renewcommand{\arraystretch}{1}
\begin{table}[h]
\caption{Permutations resulting from applying one TDRL operation
         on the identity permutation $ \pid(5) $, and the corresponding
         binary patterns that define the applied TDRL operations.}
\centering
 \begin{tabular}{ c c c c c }
 	                     &                   & &  $1 \ 2 \ 3 \ 4 \ 5$  &  \quad $(\, 1 \ 1 \ 1 \ 1 \ 1 \,)$   \\
	                     &                   & &  $1 \ 2 \ 3 \ 4 \ 5$  &  \quad $(\, 1 \ 1 \ 1 \ 1 \ 0 \,)$   \\
										   &                   & &  $1 \ 2 \ 3 \ 5 \ 4$  &  \quad $(\, 1 \ 1 \ 1 \ 0 \ 1 \,)$   \\
										   &                   & &  $1 \ 2 \ 3 \ 4 \ 5$  &  \quad $(\, 1 \ 1 \ 1 \ 0 \ 0 \,)$   \\
										   &                   & &  $1 \ 2 \ 4 \ 5 \ 3$  &  \quad $(\, 1 \ 1 \ 0 \ 1 \ 1 \,)$   \\
										   &                   & &  $1 \ 2 \ 4 \ 3 \ 5$  &  \quad $(\, 1 \ 1 \ 0 \ 1 \ 0 \,)$   \\
									  	 &                   & &  $1 \ 2 \ 5 \ 3 \ 4$  &  \quad $(\, 1 \ 1 \ 0 \ 0 \ 1 \,)$   \\
					  					 &                   & &  $1 \ 2 \ 3 \ 4 \ 5$  &  \quad $(\, 1 \ 1 \ 0 \ 0 \ 0 \,)$   \\
				  						 &                   & &  $1 \ 3 \ 4 \ 5 \ 2$  &  \quad $(\, 1 \ 0 \ 1 \ 1 \ 1 \,)$   \\
										   &    $\nearrow$     & &  $1 \ 3 \ 4 \ 2 \ 5$  &  \quad $(\, 1 \ 0 \ 1 \ 1 \ 0 \,)$   \\
										   &                   & &  $1 \ 3 \ 5 \ 2 \ 4$  &  \quad $(\, 1 \ 0 \ 1 \ 0 \ 1 \,)$   \\
										   &      $\cdot$      & &  $1 \ 3 \ 2 \ 4 \ 5$  &  \quad $(\, 1 \ 0 \ 1 \ 0 \ 0 \,)$   \\
										   &      $\cdot$      & &  $1 \ 4 \ 5 \ 2 \ 3$  &  \quad $(\, 1 \ 0 \ 0 \ 1 \ 1 \,)$   \\
				  						 &      $\cdot$      & &  $1 \ 4 \ 2 \ 3 \ 5$  &  \quad $(\, 1 \ 0 \ 0 \ 1 \ 0 \,)$   \\
                       &                   & &  $1 \ 5 \ 2 \ 3 \ 4$  &  \quad $(\, 1 \ 0 \ 0 \ 0 \ 1 \,)$   \\
	$1 \ 2 \ 3 \ 4 \ 5$  & $\longrightarrow$ & &  $1 \ 2 \ 3 \ 4 \ 5$  &  \quad $(\, 1 \ 0 \ 0 \ 0 \ 0 \,)$   \\
										   &                   & &  $2 \ 3 \ 4 \ 5 \ 1$  &  \quad $(\, 0 \ 1 \ 1 \ 1 \ 1 \,)$   \\
										   &      $\cdot$      & &  $2 \ 3 \ 4 \ 1 \ 5$  &  \quad $(\, 0 \ 1 \ 1 \ 1 \ 0 \,)$   \\
										   &      $\cdot$      & &  $2 \ 3 \ 5 \ 1 \ 4$  &  \quad $(\, 0 \ 1 \ 1 \ 0 \ 1 \,)$   \\
										   &      $\cdot$      & &  $2 \ 3 \ 1 \ 4 \ 5$  &  \quad $(\, 0 \ 1 \ 1 \ 0 \ 0 \,)$   \\
										   &                   & &  $2 \ 4 \ 5 \ 1 \ 3$  &  \quad $(\, 0 \ 1 \ 0 \ 1 \ 1 \,)$   \\
										   &    $\searrow$     & &  $2 \ 4 \ 1 \ 3 \ 5$  &  \quad $(\, 0 \ 1 \ 0 \ 1 \ 0 \,)$   \\
										   &                   & &  $2 \ 5 \ 1 \ 3 \ 4$  &  \quad $(\, 0 \ 1 \ 0 \ 0 \ 1 \,)$   \\
										   &                   & &  $2 \ 1 \ 3 \ 4 \ 5$  &  \quad $(\, 0 \ 1 \ 0 \ 0 \ 0 \,)$   \\
										   &                   & &  $3 \ 4 \ 5 \ 1 \ 2$  &  \quad $(\, 0 \ 0 \ 1 \ 1 \ 1 \,)$   \\
										   &                   & &  $3 \ 4 \ 1 \ 2 \ 5$  &  \quad $(\, 0 \ 0 \ 1 \ 1 \ 0 \,)$   \\
										   &                   & &  $3 \ 5 \ 1 \ 2 \ 4$  &  \quad $(\, 0 \ 0 \ 1 \ 0 \ 1 \,)$   \\
										   &                   & &  $3 \ 1 \ 2 \ 4 \ 5$  &  \quad $(\, 0 \ 0 \ 1 \ 0 \ 0 \,)$   \\
										   &                   & &  $4 \ 5 \ 1 \ 2 \ 3$  &  \quad $(\, 0 \ 0 \ 0 \ 1 \ 1 \,)$   \\
										   &                   & &  $4 \ 1 \ 2 \ 3 \ 5$  &  \quad $(\, 0 \ 0 \ 0 \ 1 \ 0 \,)$   \\
										   &                   & &  $5 \ 1 \ 2 \ 3 \ 4$  &  \quad $(\, 0 \ 0 \ 0 \ 0 \ 1 \,)$   \\
										   &                   & &  $1 \ 2 \ 3 \ 4 \ 5$  &  \quad $(\, 0 \ 0 \ 0 \ 0 \ 0 \,)$   
 \end{tabular}
\label{tab:tdrl}
\end{table}
}

\subsection{Counting TDRL Operations}

Define
\begin{equation}
\label{eq:S}
\begin{aligned}
  \Sout(n)  &\defeq  \big| \SSout(\pi) \big| , \quad  \Sin(n)  \defeq  \big| \SSin(\pi)  \big| ,  \\
	\Srev(n)  &\defeq  \big| \SSout(\pi) \cap \SSin(\pi) \big| .
\end{aligned}
\end{equation}
$ \Srev(n) $ can be thought of as the number of ``reversible'' TDRL operations --
those TDRL operations that can be inverted by another TDRL operation.
We first verify that the quantities $ \Sout(n) $, $ \Sin(n) $, $ \Srev(n) $ are
well-defined in that they do not depend on $ \pi $.

\begin{lemma}
\label{thm:outin}
For all $ n $ and $ \pi, \pi' \in \Pi(n) $,
$ \big| \SSout(\pi) \big| = \big| \SSout(\pi') \big| $,
$ \big| \SSin(\pi) \big| = \big| \SSin(\pi') \big| $,
$ \big| \SSout(\pi) \cap \SSin(\pi) \big| = \big| \SSout(\pi') \cap \SSin(\pi') \big| $.
\end{lemma}
\begin{IEEEproof}
A bijection between, e.g., $ \SSout(\pi) $ and $ \SSout(\pi') $, is constructed
simply by relabeling the symbols in $ \{1, 2, \ldots, n\} $ in such a way that
$ \pi $ is transformed into $ \pi' $.
More precisely, take $ \sigma \in \Pi(n) $ such that $ \sigma \circ \pi = \pi' $,
and notice that $ \pi \to \rho $ if and only if $ \sigma \circ \pi \to \sigma \circ \rho $.
\end{IEEEproof}
\vspace{1mm}



\begin{theorem}
\label{thm:Sout}
$ \Sout(n) = \Sin(n) = 2^n - n $.
\end{theorem}
\begin{IEEEproof}
Since the sequences in $ \SSout(\pid) $ are determined by binary patterns
of length $ n $, the inequality $ \big|\SSout(\pid)\big| \leq 2^n $ is
straightforward.
However, notice that the binary patterns of the form $ 1^r 0^{n-r} $,
$ 0 \leq r \leq n $, all produce the same sequence---$ \pid $ itself---so
we in fact have $ \big|\SSout(\pid)\big| \leq 2^n - n $.
To demonstrate that this upper bound is tight, one would need to show that
all other binary patterns produce different sequences.
This fact is rather obvious (see Example~\ref{ex:tdrl}) so we omit a formal
proof.

Even though the fact that $ \Sout(n) = \Sin(n) $ follows from the same
relabeling argument used in the proof of Lemma \ref{thm:outin}, we give
here an alternative derivation that is useful for understanding the
structure of reverse TDRL operations.
It follows from the definition of TDRL operations that the sequences that can
produce $ \pid $ are those that can be partitioned into \emph{subsequences}
$ (1, 2, \ldots, j) $ and $ (j+1, \ldots, n) $, for some $ j \in \{0, 1, \ldots, n \} $.
For $ j = 0, 1, \ldots, n-1 $, there are exactly $ \binom{n}{j} - 1 $ sequences
that can be partitioned into subsequences $ (1, 2, \ldots, j) $, $ (j+1, \ldots, n) $,
but cannot be partitioned into subsequences $ (1, 2, \ldots, j+1) $, $ (j+2, \ldots, n) $
(the latter condition is needed to avoid double-counting).
Namely, the number of sequences that can be partitioned into subsequences
$ (1, 2, \ldots, j) $, $ (j+1, \ldots, n) $ is the number of ways to choose
the positions for the elements of the subsequence $ (1, 2, \ldots, j) $,
which is $ \binom{n}{j} $, and among those sequences there is only one, $ \pid $,
which can also be partitioned into $ (1, 2, \ldots, j+1) $, $ (j+2, \ldots, n) $.
Therefore,
$ \big| \SSin(\pid) \big| = 1 + \sum_{j=0}^{n-1} \big( \binom{n}{j} - 1 \big) = 2^n - n $.
\end{IEEEproof}
\vspace{1mm}

We note that the identity $ \Sout(n) = 2^n - n $ also easily follows from
\cite[Thm 1.1]{chaudhuri} and \cite[Thm 6]{bouvel+rossin}.

In the following statement we obtain an expression for the number of reversible
TDRL operations, or equivalently, for the number of sequences that can both
produce $ \pid $ and be produced by it.

\begin{theorem}
\label{thm:Srev}
$ \Srev(n) = 1 + \binom{n}{2} + \binom{n}{3} $.
\end{theorem}
\begin{IEEEproof}
We first argue that a TDRL operation is reversible if and only if
the corresponding binary pattern is of the form $ b = 1^r 0^s 1^t 0^u $, where
$ r, s, t, u $ are non-negative integers summing to $ n $.
In words, the requirement is that $ b $ has at most two blocks of ones, and if
it has exactly two blocks, then one of them is the leading block.
For the direct part, notice that a TDRL operation $ 1^r 0^s 1^t 0^u $ is
reversible by the TDRL operation $ 1^r 0^t 1^s 0^u $.
Conversely, if a TDRL operation is not of the form $ 1^r 0^s 1^t 0^u $, then
its binary pattern can be written as $ a\, 0^r 1^s 0^t 1^u b $, where $ r, s, t, u $
are strictly positive integers and $ a $ and $ b $ are arbitrary (possibly
empty) binary strings.
Such a TDRL operation produces a sequence that cannot be partitioned into
subsequences $ (1, 2, \ldots, j) $, $ (j+1, \ldots, n) $ and is therefore not
reversible.

Now that we have a characterization of reversible TDRL operations, we can use
it to show the desired expression.
There is one binary pattern containing no $ 1 $'s, and there are $ \binom{n + 1}{2} $
binary patterns containing exactly one block of $ 1 $'s (a block is determined
by its delimiters).
Among the latter, there are $ n $ patterns for which this block is the leading
block, i.e., patterns of the form $ 1^r 0^{n-r} $, $ r > 0 $.
As we already know, such patterns correspond to the same TDRL operation as the
pattern $ 0 0 \cdots 0 $, while all the other patterns correspond to different
TDRL operations.
Therefore, there are exactly $ 1 + \binom{n+1}{2} - n = 1 + \binom{n}{2} $ different
TDRL operations corresponding to binary patterns with at most one block of $ 1 $'s.
Finally, there are $ \binom{n}{3} $ binary patterns with exactly two blocks of
$ 1 $'s, one of which is the leading block (choose the length of the leading
block and then choose the delimiters of the second block), and all of them
correspond to different TDRL operations.
\end{IEEEproof}
\vspace{1mm}

Thus, only an asymptotically vanishing fraction of TDRL operations are
reversible, $ \lim_{n \to \infty} \Srev(n) / \Sout(n) = 0 $.

\subsection{The Reconstruction Problem}
\label{sec:rec}

The sequence reconstruction problem, as introduced by Levenshtein \cite{levenshtein_IT},
is defined as follows: a sequence $ \bs{x} $ is transmitted through a noisy channel
multiple times, and the receiver is required to reconstruct it after it has collected
sufficiently many noisy observations.
The question is how many different noisy versions of the sequence are sufficient
in order to guarantee successful and unambiguous reconstruction.
In combinatorial terms the problem can be rephrased as follows: what is the cardinality
of the largest possible intersection of sets of channel outputs that two different
sequences of length $ n $ can produce?
Denoting the cardinality of the mentioned largest intersection by $ N(n) $, one
easily concludes that the number of noisy observations that guarantees successful
reconstruction in all cases is $ N(n) + 1 $.
The problem of determining the largest intersection of two ``balls'' in a given
space is therefore relevant in all situations where one uses a simple repetition
scheme to communicate reliably.
As argued in \cite{yehezkeally+schwartz}, this problem naturally arises in DNA-based
data storage applications.

In the present context, the ``noise'' are the TDRL rearrangement operations and
the reconstruction problem reduces to the following:
what is the largest possible cardinality of the set $ \SSout(\pi) \cap \SSout(\rho) $?
So define
\begin{equation}
  N(n)  \defeq  \max_{\substack{\pi, \rho \,\in\, \Pi(n) \\ \pi \neq \rho}}
	                   \big| \SSout(\pi) \cap \SSout(\rho) \big| .
\end{equation}
In the following statement we give a solution to the reconstruction problem
just described.
For other relevant works on the reconstruction problem for translocation/permutation
errors, see, e.g., \cite{konstantinova, levenshtein+siemons, yaakobi}.

\begin{theorem}
\label{thm:N}
$ N(n) = 2^{n-1} $.
\end{theorem}
\begin{IEEEproof}
Consider the sequence $ \pi = (2, 3, \ldots, n, 1) $ obtained from $ \pid $
by moving the first symbol to the last position (a cyclic shift).
Consider some $ \rho \in \SSout(\pi) $, and suppose that the binary pattern
corresponding to the TDRL operation $ \pi \to \rho $ ends in a $ 1 $, i.e.,
is of the form $ b\,1 $ for $ b \in \{0, 1\}^{n-1} $.
Then it is easy to see that $ \rho $ can also be obtained from $ \pid $ via
the TDRL operation $ 0\,b $, and hence $ \rho \in \SSout(\pid) $.
Since there are $ 2^{n-1} $ binary strings of the form $ b\,1 $, and since
all of them result in different sequences $ \rho $, we have just shown that
$ \big| \SSout(\pid) \cap \SSout(\pi) \big| \geq 2^{n-1} $, and therefore
$ N(n) \geq 2^{n-1} $.

We now use induction to prove that $ \big| \SSout(\pid) \cap \SSout(\pi) \big| \leq 2^{n-1} $
for every $ n \geq 2 $ and every $ \pi \in \Pi(n) \setminus \{\pid\} $.
Suppose that, for a given $ n \geq 3 $, there is a sequence $ \pi = (\pi_1, \pi_2, \ldots, \pi_n) \in \Pi(n) $
such that $ \big| \SSout(\pid) \cap \SSout(\pi) \big| > 2^{n-1} $.
This implies that there are at least $ 2^{n-1} + 1 $ binary patterns describing
TDRL operations $ \pi \to \rho $ such that $ \rho \in \SSout(\pid) \cap \SSout(\pi) $.
If $ \pi_i = j $, denote $ \pi_{\setminus i} = (\pi_1, \ldots, \pi_{i-1}, \pi_{i+1}, \ldots, \pi_n) $
and $ \sigma_{\setminus i} = (1, \ldots, j-1, j+1, \ldots, n) $, and suppose
that $ \pi_{\setminus i} \neq \sigma_{\setminus i} $ (if not, choose another
index $ i $ for which this holds).
(By possibly renaming the symbols, both $ \pi_{\setminus i} $ and $ \sigma_{\setminus i} $
can be thought of as sequences/permutations over $ \{1, 2, \ldots, n-1\} $, in
which case $ \sigma_{\setminus i} $ would be the identity permutation.)
By deleting the $ i $'th bit of each of the mentioned binary patterns, one would
get at least $ 2^{n-2} + 1 $ different binary patterns of length $ n-1 $.
Notice that these binary patterns describe TDRL operations on the sequence
$ \pi_{\setminus i} $, and that every sequence $ \rho' $ that is the result of
such an operation can be produced by $ \sigma_{\setminus i} $ as well, i.e.,
$ \rho' \in \SSout(\sigma_{\setminus i}) \cap \SSout(\pi_{\setminus i}) $.
(If a binary pattern $ b $ describes a TDRL operation $ \pi \to \rho $ that
produces a sequence $ \rho $ in the intersection $ \SSout(\pid) \cap \SSout(\pi) $,
then it is not difficult to see that the pattern $ b_{\setminus i} $ describes
a TDRL operation $ \pi_{\setminus i} \to \rho' $ that produces a sequence $ \rho' $
in the intersection $ \SSout(\sigma_{\setminus i}) \cap \SSout(\pi_{\setminus i}) $.)
We have thus shown that the assumption $ N(n) > 2^{n-1} $ implies that $ N(n-1) > 2^{n-2} $.
In other words, assuming $ N(n-1) \leq 2^{n-2} $ implies $ N(n) \leq 2^{n-1} $,
and since one can directly verify that $ N(2) = 2 $, the inductive proof that
$ N(n) \leq 2^{n-1} $ for every $ n $ is complete.
\end{IEEEproof}
\vspace{1mm}

As exemplified in the previous proof, the intersection $ \SSout(\pi) \cap \SSout(\rho) $
is of maximum possible cardinality when $ \pi, \rho $ are cyclic shifts (by one
position) of one another.
This is also the case for any two sequences $ \pi, \rho $ that differ by one
adjacent transposition, e.g.,
$ \pi = \pid = (1, 2, 3, \ldots, n) $, $ \rho = (2, 1, 3, \ldots, n) $.

\begin{corollary}
Let $ n \geq 3 $.
Every sequence $ \pi \in \Pi(n) $ is uniquely determined by any $ 2^{n-1} + 1 $
elements of $ \SSout(\pi) $.
\end{corollary}
\begin{IEEEproof}
We just have to verify that
$ \big|\SSout(\pi)\big| = 2^n - n \geq 2^{n-1} + 1 = N(n) + 1 $ for $ n \geq 3 $.
\end{IEEEproof}

\subsection{Bounded TDRL Permutations}
\label{sec:bounded}

In this subsection we analyze a more general model where a TDRL rearrangement
operation is confined to segments of width $ k $ within the original sequence
\cite{bouvel+rossin}.
In other words, a TDRL operation is in this case applied on a segment of $ k $
consecutive symbols of a given sequence $ \pi $, while the remaining symbols of
$ \pi $ are left intact.

\begin{example}
\label{ex:tdrlnk}
One possible TDRL operation on $ \pid(5) $, applied on the segment $ (2, 3, 4) $
of length $ k = 3 $, is the following:
\begin{align}
\label{eq:tdrlnk5}
	1 \ \und{2} \ 3 \ \und{4} \ \aoverbrace[L1R]{2 \ \und{3} \ 4} \ 5 
	\quad \longrightarrow \quad  1 \ 2 \ 4 \ 3 \ 5  
\end{align}
where the duplicate segment is overbraced, and the symbols that are \emph{not}
deleted (from the original segment $ (2, 3, 4) $ and its duplicate) are underlined.
\myqed
\end{example}

In the special case $ k = 2 $, the only non-trivial TDRL operations are
adjacent transpositions, i.e., swaps of two adjacent symbols.

Let $ \Sout(n; k) $ be the number of sequences that can be obtained from $ \pi \in \Pi(n) $
by applying a TDRL operation on an arbitrary segment of $ \pi $ consisting of $ k $
consecutive symbols, and define $ \Sin(n; k) $ and $ \Srev(n; k) $ accordingly
(see~\eqref{eq:S}).
The same argument that was used in the proof of Lemma~\ref{thm:outin} can be
used in this context as well, implying that $ \Sin(n; k) = \Sout(n; k) $.

\begin{theorem}
\label{thm:Soutnk}
$ \Sout(n; k) = \Sin(n; k) = (n - k + 2)(2^{k-1} - 1) - k + 2 $.
\end{theorem}
\begin{IEEEproof}
Consider first the sequences that can be produced from $ \pid(n) $ by applying
a TDRL operation on its first $ k $ symbols.
We know by Theorem \ref{thm:Sout} that the number of such sequences is $ 2^k - k $.
Now consider the second ``window'' of length $ k $ containing the symbols
$ 2, 3, \ldots, k+1 $.
There are again $ 2^k - k $ different sequences we can get by applying a TDRL
operation on this window; however, some of them are identical to sequences
that were obtained in the first step.
Namely, all sequences that can be produced by a TDRL operation on the intersection
of the two windows, i.e., on the symbols $ 2, 3, \ldots, k $, are double-counted
in this way,
\[
  \rlap{$\overbracket{\phantom{1\ 2\ 3\ 4}}$}  1\, \underbracket{2\ 3\ 4\ 5}\,  6\ 7 .
\]
The number of sequences that have been double-counted---those that can be produced
by a TDRL operation on the segment $ 2, 3, \ldots, k $---is 
$ 2^{k-1} - (k-1) $.
We then proceed to find $ \Sout(n; k) $ as follows: count the sequences that can
be produced by a TDRL operation on $ 1, 2, \ldots, k $ but cannot be produced by
a TDRL operation on $ 2, \ldots, k $ (the latter will be counted in the second window);
then add the number of sequences that can be produced by a TDRL operation on
$ 2, 3, \ldots, k+1 $ but cannot be produced by a TDRL operation on $ 3, \ldots, k+1 $;
etc.
This is done for the first $ n-k $ windows.
For the last, $ (n-k+1) $'th window there is no need exclude any sequences because
the procedure stops and there is no double-counting.
We thus get $ \Sout(n; k) = (n-k)(2^k-k-(2^{k-1} - (k-1))) + 2^k - k $, which is
what we needed to show.
\end{IEEEproof}
\vspace{1mm}

As an application of Theorem \ref{thm:Soutnk}, we next state a sphere-packing
bound for codes in $ \Pi(n) $ correcting one ``TDRL error'' of length $ k $.
Namely, let $ \bs{C} \subseteq\Pi(n) $ be a set of sequences with the property
that every sequence from $ \bs{C} $ can be uniquely recovered even after a TDRL
operation of length $ k $ has been applied on it.
Then, by Theorem \ref{thm:Soutnk} and a simple sphere-packing argument, we
conclude that the cardinality of any such code is upper-bounded as:
\begin{equation}
\label{eq:sp}
  |\bs{C}| \leq \frac{ | \Pi(n) | }{ \Sout(n; k) } = \frac{ n! }{ (n - k + 2)(2^{k-1} - 1) - k + 2 } .
\end{equation}
For $ k = 2 $ we have $ \Sout(n; k) = n $, and the above sphere-packing bound
reduces to $ |\bs{C}| \leq (n-1)! $.
We note that error-correcting codes in $ \Pi(n) $ with respect to various
error/rearrangement models have been extensively studied in the literature;
see, e.g., \cite{barg+mazumdar, buzaglo+etzion, farnoud, gabrys} and the
references therein.

\begin{theorem}
\label{thm:Srevnk}
$ \Srev(n; k) = (n-k+1)\binom{k}{2} + \binom{k}{3} + 1 $.
\end{theorem}
\begin{IEEEproof}
The statement follows from Theorem \ref{thm:Srev} and the inclusion-exclusion
method of counting that was used in the proof of Theorem \ref{thm:Soutnk} as well.
\end{IEEEproof}
\vspace{1mm}

Note that Theorems \ref{thm:Sout}, \ref{thm:Srev} are recovered from Theorems
\ref{thm:Soutnk}, \ref{thm:Srevnk} for $ n = k $.

\section{Mirror-TDRL Permutations}

A mirror (or palindromic) TDRL operation---MTDRL operation for short---on a
sequence $ \pi \in \Pi(n) $ is a duplication of the sequence $ \pi $, followed
by a \emph{reversal} of the second copy, and by a deletion of one of the two
copies of each of the individual symbols \cite{baril+vernay}.

\begin{example}
\label{ex:mtdrl}
An example of a MTDRL operation on $ \pid(5) $ is the following:
\begin{subequations}
\label{eq:mtdrl}
\begin{align}
\label{eq:mtdrl5}
  &1 \ \und{2} \ \und{3} \ 4 \ \und{5}  \ \aoverbrace[L1R]{ 5 \ \und{4} \ 3 \ 2 \ \und{1} }
  \quad \longrightarrow \quad  2 \ 3 \ 5 \ 4 \ 1  \\
  &0 \ 1 \ 1 \ 0 \ 1
\end{align}
\end{subequations}
where the reversed copy of the original sequence is overbraced, and the symbols
that are \emph{not} deleted are underlined.
\myqed
\end{example}

The set of sequences resulting from applying a MTDRL operation on $ \pid $
is illustrated in Table \ref{tab:mtdrl}.

{
\renewcommand{\arraystretch}{1.05}
\begin{table} 
\caption{Permutations resulting from applying one MTDRL operation
         on the identity permutation $ \pid(4) $, and the
				 corresponding binary patterns that define the applied MTDRL operations.}
\centering
 \begin{tabular}{ c c c c c }
 	                     &                   & &  $1 \ 2 \ 3 \ 4$  &  \quad $(\, 1 \ 1 \ 1 \ 1 \,)$   \\
	                     &                   & &  $1 \ 2 \ 3 \ 4$  &  \quad $(\, 1 \ 1 \ 1 \ 0 \,)$   \\
										   &                   & &  $1 \ 2 \ 4 \ 3$  &  \quad $(\, 1 \ 1 \ 0 \ 1 \,)$   \\
										   &    $\nearrow$     & &  $1 \ 2 \ 4 \ 3$  &  \quad $(\, 1 \ 1 \ 0 \ 0 \,)$   \\
										   &      $\cdot$      & &  $1 \ 3 \ 4 \ 2$  &  \quad $(\, 1 \ 0 \ 1 \ 1 \,)$   \\
										   &      $\cdot$      & &  $1 \ 3 \ 4 \ 2$  &  \quad $(\, 1 \ 0 \ 1 \ 0 \,)$   \\
									  	 &      $\cdot$      & &  $1 \ 4 \ 3 \ 2$  &  \quad $(\, 1 \ 0 \ 0 \ 1 \,)$   \\
		$1 \ 2 \ 3 \ 4$		 & $\longrightarrow$ & &  $1 \ 4 \ 3 \ 2$  &  \quad $(\, 1 \ 0 \ 0 \ 0 \,)$   \\
				  						 &      $\cdot$      & &  $2 \ 3 \ 4 \ 1$  &  \quad $(\, 0 \ 1 \ 1 \ 1 \,)$   \\
										   &      $\cdot$      & &  $2 \ 3 \ 4 \ 1$  &  \quad $(\, 0 \ 1 \ 1 \ 0 \,)$   \\
										   &      $\cdot$      & &  $2 \ 4 \ 3 \ 1$  &  \quad $(\, 0 \ 1 \ 0 \ 1 \,)$   \\
										   &    $\searrow$     & &  $2 \ 4 \ 3 \ 1$  &  \quad $(\, 0 \ 1 \ 0 \ 0 \,)$   \\
										   &                   & &  $3 \ 4 \ 2 \ 1$  &  \quad $(\, 0 \ 0 \ 1 \ 1 \,)$   \\
				  						 &                   & &  $3 \ 4 \ 2 \ 1$  &  \quad $(\, 0 \ 0 \ 1 \ 0 \,)$   \\
                       &                   & &  $4 \ 3 \ 2 \ 1$  &  \quad $(\, 0 \ 0 \ 0 \ 1 \,)$   \\
                       &                   & &  $4 \ 3 \ 2 \ 1$  &  \quad $(\, 0 \ 0 \ 0 \ 0 \,)$   
 \end{tabular}
\label{tab:mtdrl}
\end{table}
}

\subsection{Counting MTDRL Operations}

The quantities $ \Smout(n) $, $ \Smin(n) $, $ \Smrev(n) $ in this setting are
defined similarly to \eqref{eq:S}.
The fact that $ \Smout(n) = \Smin(n) $ is established by the same reasoning as
in Lemma \ref{thm:outin}.

\begin{theorem}
\label{thm:Smout}
$ \Smout(n) = \Smin(n) = 2^{n-1} $.
\end{theorem}
\begin{IEEEproof}
The binary patterns $ b\,1 $ and $ b\,0 $, for $ b \in \{0,1\}^{n-1} $, always
produce the same sequence.
This follows from the definition of MTDRL operations \eqref{eq:mtdrl} (see also
Table \ref{tab:mtdrl}).
Furthermore, all patterns $ b\,1 $, $ b \in \{0,1\}^{n-1} $, produce different
sequences.
Hence, $ \Smout(n) = 2^{n-1} $.
\end{IEEEproof}
\vspace{1mm}

We note that Theorem \ref{thm:Smout} can also be inferred from the characterization
of the set of sequences $ \SSmout(\pid) $ obtained in \cite[Lem.~2 and Cor.~1]{baril+vernay}.
The following statement gives the number of reversible MTDRL operations.

\begin{theorem}
\label{thm:Smrev}
$ \Smrev(n) = n $.
\end{theorem}
\begin{IEEEproof}
We need to count all sequences that can both produce $ \pid $ and be produced
by it in a single MTDRL operation.
First notice that all sequences in $ \SSmout(\pid) $ are unimodular (first
increasing, then decreasing).
This follows from the definition of MTDRL operations -- each such operation can
be seen as selecting a (necessarily increasing) subsequence of $ \pid $ in the
first step, and then reading off the remaining subsequence in reverse order.
Now, if a sequence $ \rho \in \SSmout(\pid) $ ends with $ 1 $, it is possible
to produce $ \pid $ from it only via the pattern $ 0 \cdots 0\,1 $ (because
$ \pid $ starts with $ 1 $), which implies that $ \rho = (n, n-1, \ldots, 2, 1) $.
If a sequence $ \rho \in \SSmout(\pid) $ ends with $ 2 $, then it has to start
with $ 1 $ because it is unimodular, as we have noted above.
It is possible to produce $ \pid $ from such a sequence only via the pattern
$ 1\,0 \cdots 0\,1 $ (because $ \pid $ starts with $ 1,2 $), which implies that
$ \rho = (1, n, n-1, \ldots, 3, 2) $.
Continuing in this way, one concludes that there is exactly one sequence
$ \rho \in \SSmout(\pid) $ that ends with $ i $, $ i \in \{1, 2, \ldots, n\} $,
and that can produce $ \pid $.
Therefore, the number of reversible MTDRL operations is $ n $.
\end{IEEEproof}

\subsection{The Reconstruction Problem}

We next determine the maximum cardinality of the intersections
$ \SSmout(\pi) \cap \SSmout(\rho) $, pertaining to the reconstruction problem
as defined in Section \ref{sec:rec}.
Let
\begin{equation}
\label{eq:Nm}
  N_\textsc{m}(n)  \defeq  \max_{\substack{\pi, \rho \,\in\, \Pi(n) \\ \pi \neq \rho}}
	                               \big| \SSmout(\pi) \cap \SSmout(\rho) \big| .
\end{equation}
As it turns out, $ N_\textsc{m}(n) = \Smout(n) $ for all $ n $, and therefore
unambiguous reconstruction is in general impossible for MTDRL operations.

\begin{theorem}
\label{thm:Nm}
$ N_\textsc{m}(n) = 2^{n-1} $.
\end{theorem}
\begin{IEEEproof}
Consider the sequence $ \pi = (1, 2, \ldots, n-2, n,\linebreak n-1) $ obtained
from $ \pid $ by swapping its last two elements.
Recall that every sequence in $ \SSmout(\pid) $ can be obtained from $ \pid $
via a MTDRL operation whose binary pattern is of the form $ b\,1 $, $ b \in \{0,1\}^{n-1} $.
Furthermore, it can be easily checked that $ \pid \to \rho $ via $ b\,0\,1 $ if
and only if $ \pi \to \rho $ via $ b\,1\,1 $, where $ b \in \{0,1\}^{n-2} $.
Likewise, $ \pid \to \rho $ via $ b\,1\,1 $ if and only if $ \pi \to \rho $ via $ b\,0\,1 $.
This shows that every $ \rho $ that belongs to $ \SSmout(\pid) $ also belongs
to $ \SSmout(\pi) $, and thus
$ \big| \SSmout(\pid) \cap \SSmout(\pi) \big| = \big| \SSmout(\pid) \big| = 2^{n-1} $.
\end{IEEEproof}

\subsection{Bounded MTDRL Permutations}

Consider now a more general model where MTDRL rearrangement operations are
confined to segments of width $ k $ within the original sequence (see Section~\ref{sec:bounded}),
and define $ \Smout(n; k) $, $ \Smin(n; k) $, $ \Smrev(n; k) $ accordingly.

\begin{theorem}
\label{thm:Smoutnk}
$ \Smout(n; k) = \Smin(n; k) = (n - k + 1)(2^{k-1} - 1) + 1 $.
\end{theorem}
\begin{IEEEproof}
We use the inclusion-exclusion counting method for ``sliding window'' of width
$ k $, as for the TDRL model (see the proof of Theorem \ref{thm:Soutnk}).
The main question is how many sequences need to be excluded for a given window
in order to avoid double-counting?
It turns out that the situation for {MTDRL} is simpler than for TDRL, and only
one sequence needs to excluded -- the identity permutation.
Namely, any non-trivial MTDRL operation on the window $ (2, 3, \ldots, k+1) $
results in the last symbol ($ k+1 $) being moved to one the preceding positions
(see \eqref{eq:mtdrl}), and applying a MTDRL operation to the window $ (1, 2, \ldots, k) $
clearly leaves the symbol $ k+1 $ intact.
Therefore, only one sequence---$ \pid(n) $ itself---can be produced by both a
MTDRL operation on the segment $ (1, 2, \ldots, k) $ and a MTDRL operation on
the segment $ (2, 3, \ldots, k+1) $ of $ \pid(n)) $.
By using this fact and Theorem~\ref{thm:Smout}, we get
$ \Smout(n; k) = (n - k)(2^{k-1} - 1) + 2^{k-1} $.
\end{IEEEproof}
\vspace{1mm}

If $ \bs{C}_\textsc{m} \subseteq \Pi(n) $ is a code that is able to recover
from one MTDRL operation of length $ k $, then, by Theorem \ref{thm:Smoutnk}
and a simple sphere-packing argument, we obtain the following bound on its
cardinality:
\begin{equation}
\label{eq:msp}
  |\bs{C}_\textsc{m}| \leq \frac{ | \Pi(n) | }{ \Smout(n; k) } = \frac{ n! }{ (n - k + 1)(2^{k-1} - 1) + 1 } .
\end{equation}

\pagebreak
\begin{theorem}
\label{thm:Smrevnk}
$ \Smrev(n; k) = (n-k+1)(k-1) + 1 $.
\end{theorem}
\begin{IEEEproof}
Follows from Theorem \ref{thm:Smrev} after applying the same method of
counting as in the proof of Theorem \ref{thm:Smoutnk}.
\end{IEEEproof}
\vspace{2mm}

\subsubsection*{Acknowledgment}

This work was supported by the European Commission (H2020 Antares project, ref. no. 739570).


\end{document}